# REX: Explaining Relationships between Entity Pairs[*]

Lujun Fang[†], Anish Das Sarma[‡], Cong Yu[‡], Philip Bohannon[♯]
[†]University of Michigan, [♯]Yahoo! Research, [‡]Google Research
ljfang@umich.edu, {anish,congyu}@google.com, plb@yahoo-inc.com

## ABSTRACT

Knowledge bases of entities and relations (either constructed manually or automatically) are behind many real world search engines, including those at Yahoo!, Microsoft[1], and Google. Those knowledge bases can be viewed as graphs with nodes representing entities and edges representing (primary) relationships, and various studies have been conducted on how to leverage them to answer entity seeking queries. Meanwhile, in a complementary direction, analyses over the query logs have enabled researchers to identify entity pairs that are statistically correlated. Such entity relationships are then presented to search users through the "related searches" feature in modern search engines. However, entity relationships thus discovered can often be "puzzling" to the users because why the entities are connected is often indescribable. In this paper, we propose a novel problem called *entity relationship explanation*, which seeks to explain why a pair of entities are connected, and solve this challenging problem by integrating the above two complementary approaches, i.e., we leverage the knowledge base to "explain" the connections discovered between entity pairs.

More specifically, we present *REX*, a system that takes a pair of entities in a given knowledge base as input and efficiently identifies a ranked list of relationship explanations. We formally define relationship explanations and analyze their desirable properties. Furthermore, we design and implement algorithms to efficiently enumerate and rank all relationship explanations based on multiple measures of "interestingness." We perform extensive experiments over real web-scale data gathered from DBpedia and a commercial search engine, demonstrating the efficiency and scalability of *REX*. We also perform user studies to corroborate the effectiveness of explanations generated by *REX*.

## 1. INTRODUCTION

Search companies have been eager to evolve beyond the "*ten blue links*" model and are introducing a suite of features to help online users search and explore information more effectively. Among those features, one of the most intuitive is the so called *related entities*: when a user searches for an entity, a list of entities that are *in some way* related to the given entity are also shown to the user. This feature can be seen on major search engines like Google and Yahoo! (screenshots in Figure 1).

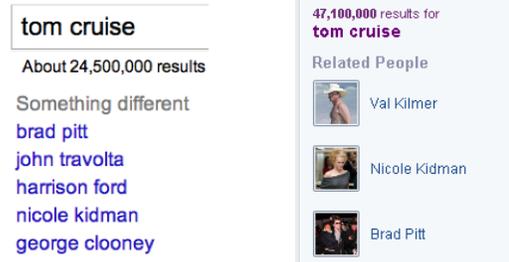

**Figure 1: Related entities feature on left panel of Google (left) and Yahoo! (right).**

However, given an entity, why certain entities are considered related is often a mystery to the user. For example, it is difficult for users other than film junkies to understand why 'Tom Cruise' and 'Brad Pitt' are related, beyond the fact that they are both popular actors. Informal user studies at Yahoo! indicate that augmenting related suggestions with concrete explanations would significantly increase the relevance of the suggestions and increase user engagement. Motivated by these studies, we aim to eliminate the mystery behind suggestions by providing *relationship explanations*: Given a pair of entities, our goal is to effectively and efficiently produce explanations that describe how the entities are related, based on a large knowledge base that maintains structured information about all entities[2]. We chose knowledge bases as the sources for explanations because of their wide spread availability behind search engines. As a very simple example of such an explanation, when 'Nicole Kidman' is shown as related to 'Tom Cruise', we would like to let the users know that they used to be married. A slightly more sophisticated explanation arises when 'Brad Pitt' is shown as related to 'Tom Cruise': we would like to show that they co-starred in a number of movies, perhaps including example(s) of such movie(s), say 'Interview with the Vampires'.

In this study, we choose to separate the explanation generation mechanism from the related entity selection mechanism, and focus on generating explanations, given a pair of entities already found to be related. The main motivation for decoupling explanations based on a knowledge graph from the reason a pair of entities was deemed related is that, in most search engines, the related entities generation mechanisms are not semantically meaningful. E.g., two entities can be considered related simply because search users often query them together in one session.

---

[*]Work partially done while Lujun Fang, Anish Das Sarma, Cong Yu were at Yahoo! Research.
[1] http://entitycube.research.microsoft.com/

[2]Note that we are separating the explanation generation mechanism from the related entity selection mechanism, and focus on generating explanations given a pair of entities already found to be related.

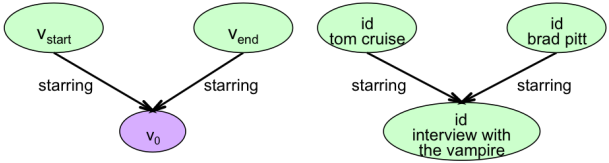

Figure 2: Example explanation for 'Tom Cruise' & 'Brad Pitt'. The graph pattern is on the left and one of the instances associated with the pattern is on the right.

Intuitively, we consider a relationship explanation as a constrained graph pattern and its associated graph instances derivable from the underlying knowledge base. Specifically, the graph pattern (similar to a graph query) contains variables as nodes and labeled relationships as edges, and the instances can be considered as the results of applying the graph pattern on the underlying knowledge base. One such example is shown in Figure 2 and we shall introduce the formal definitions in Section 2.

The overall process of relationship explanation consists of two main steps: (1) *Explanation Enumeration:* Given two entities, the starting one (i.e., the one user searched for) and the ending one (i.e., the one being suggested by the search engine), identify a list of candidate explanations; (2) *Explanation Ranking:* Rank the candidate explanations based on a set of measures to identify the most interesting explanations to be returned to the user. Both steps involve significant semantic and algorithmic challenges. First, since the knowledge base typically contains several million nodes, efficiently enumerating candidate explanations is an arduous task. Second, explanation ranking involves two significant challenges: defining suitable measures that can effectively capture explanations' interestingness and computing those measures for a large number of explanations in almost real time. Finally, we also seek opportunities to perform aggressive pruning when combining enumeration and ranking.

It is worth noting that there are quite a few existing works on mining connecting structures from graphs, such as keyword search in relational and semi-structured databases [1, 2, 3, 5, 17, 12, 13, 14, 21, 24, 29, 15] and graph mining [8, 10, 18, 22, 25]. The *key differentiating contribution of REX* is to consider connection structures that are more complex than trees and paths for explaining two entities, and introduce two novel families of pattern level interestingness measures.

To the best of our knowledge, this is the first work addressing and formalizing the problem of generating relationship explanations for a pair of entities. We make the following **main contributions**: First, we formally define the notion of *relationship explanation* and carefully analyze the properties of desirable explanations (Section 2). Second, we design and implement efficient algorithms for enumerating candidate explanations (Section 3). Third, we propose different interestingness measures for ranking relationship explanations, and design and implement efficient algorithms for ranking explanations efficiently (Section 4). Finally, we perform user studies and extensive experiments to demonstrate the effectiveness and efficiency of our algorithms (Section 5).

## 2. FUNDAMENTALS

In this section, we formally introduce the relationship explanation problem. We start by describing the input *knowledge base* (Section 2.1) from which the relationship explanations are generated. In Section 2.2, we introduce the formal definition for relation-ship explanation, which is composed of two essential components: *relationship explanation pattern* and *relationship explanation instances*. In Section 2.3, we describe important properties of relationship explanations in terms of the graph structure. The subset of relationship explanations that best satisfy the desired properties are called *minimal explanations* and are explored in the remaining of our study.

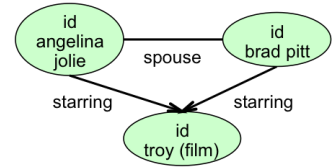

Figure 3: A subset of the entertainment knowledge base.

### 2.1 Knowledge Base

As motivated in Section 1, we choose to construct explanations from an input *knowledge base*, which is formally represented as a graph that consists of *entities* (e.g., persons, movies, etc.) as nodes, and *primary relationships* between entities (e.g., starring, spouse, etc.) as edges[3]. Entities have unique IDs (e.g., brad pitt)[4] and edges can be either directed (e.g., starring) or undirected (e.g., spouse). Therefore a knowledge base can be represented as a three-tuple $G = (V, E, \lambda)$, where $V$ is the set of nodes, $E$ is the set of edges, and $\lambda = E \to \Sigma$ is the edge labeling function.

Figure 3 illustrates a simple running example, which is a subset of the entertainment knowledge base behind the Yahoo! search engine (the actual knowledge base contains 200K nodes and over 1M edges extracted from DBPedia). The primary relationships are represented as solid lines with arrows (directed relationships) or without arrows (undirected relationships).

### 2.2 Relationship Explanation

Intuitively, a relationship explanation is a constrained graph pattern along with its associated instances that are derivable from the knowledge base. We use the terms *relationship explanation pattern* and *relationship explanation instance* to describe the two components respectively. The existence of a relationship explanation pattern is independent of the knowledge base. However, an explanation pattern is only meaningful if its associated relationship explanation instances can be found in the knowledge base with respect to the given entity pair. More concretely, the relationship explanation pattern is modeled as a graph structure that connects two target nodes representing the given entity pair. Edges in the structure have constant labels and the remaining nodes in the structure are variables:

DEFINITION 1 (RELATIONSHIP EXPLANATION PATTERN). *A relationship explanation pattern can be represented as a 5-tuple, $p = (V, E, \lambda, v_{start}, v_{end})$, where $V$ is the set of node variables, with two special variables $v_{start}$ and $v_{end}$, $E$ is a multiset of edges, and $\lambda = E \to \Sigma$ is the edge labeling function.*

Relationship explanation instances, on the other hand, capture the actual data instances from the knowledge base and are used to support an explanation pattern. Intuitively, given the knowledge base $G$, a pair of related entities that map to two nodes $v_{start}$ and $v_{end}$ in $G$, and an explanation pattern $p$, explanation instances for

---
[3] We use the term primary relationships to distinguish them from the *derived relationships* that REX will infer during the construction of the explanations.
[4] In practice, the IDs are system generated, but for the simplicity of discussion, we adopt readable titles/names as the IDs.

242

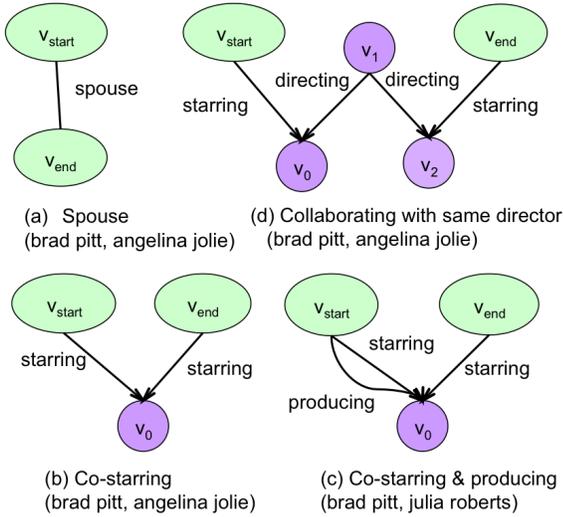

(a) Spouse
(brad pitt, angelina jolie)

(d) Collaborating with same director
(brad pitt, angelina jolie)

(b) Co-starring
(brad pitt, angelina jolie)

(c) Co-starring & producing
(brad pitt, julia roberts)

**Figure 4: Example explanation patterns.**

$p$ can be defined based on mappings from $p$ to $G$, identifying the subgraphs of $G$ that satisfy the explanation pattern.

DEFINITION 2 (RELATIONSHIP EXPLANATION INSTANCE). *Given the knowledge base $G = (V, E, \lambda)$, an explanation pattern $p = (V', E', \lambda', v'_{start}, v'_{end})$, and two target nodes $v_{start}, v_{end} \in V$, an* explanation instance *of $p$, denoted as $i(p, G, v_{start}, v_{end})$, or $i_p$, is a mapping $f : V' \rightarrow V$, where $v'_{start}$ is mapped to $v_{start}$, $v'_{end}$ is mapped to $v_{end}$ and nodes in $V' - \{v'_{start}, v'_{end}\}$ are mapped into $V - \{v_{start}, v_{end}\}$. Edge constraints must be satisfied: $\forall e' = (v'_1, v'_2) \in E'$ there must be an edge $(f(v'_1), f(v'_2))$ with label $\lambda'(e')$ in $G$. The set of all $p$'s instances are denoted as $I(p, G, v_{start}, v_{end})$, or $I_p$.*

For a pair of entities $v_{start}$ and $v_{end}$, a **relationship explanation** is defined as the pair $(p, I_p)$ consisting of the explanation pattern $p$ and the explanation instances $I_p$, where $|I_p| \geq 0$.

EXAMPLE 1. *Figure 4 illustrates some relationship explanation patterns that have at least one instance from our entertainment knowledge base between 'Brad Pitt' and 'Angelina Jolie' or 'Julia Roberts'. In particular, Figure 4(a) shows a most simple spouse relationship pattern. Figure 4(b) shows the* co-starring *relationship pattern, i.e., both 'Brad Pitt' and 'Angelina Jolie' starred together in one or more movies (which are collectively represented as the variable node $v_0$). Figures 4(c) and 4(d) illustrate more complicated relationship explanation patterns: the former adds the producing relationship between 'Brad Pitt' and the movie variable $v_0$ to produce an explanation pattern slightly more complicated than co-starring, while the latter introduces one additional movie variable ($v_2$) and one director variable ($v_1$) to form the "collaborating with same director" explanation pattern.*

### 2.3 Properties of Explanations

Definitions 1 and 2 allow a very large space of possible explanations, some of which may not be semantically meaningful. This prompted us to identify desirable structural properties of the explanations, which are described below. We note that since the structures of the instances are enforced by their corresponding patterns, we discuss the structural properties in terms of the patterns. Later, in Section 4, we describe how instances are critical in determining the interestingness of the explanations.

**Essentiality**
We want to capture the desideratum that explanation patterns contain only the "essential" nodes or edges, i.e., all nodes and edges

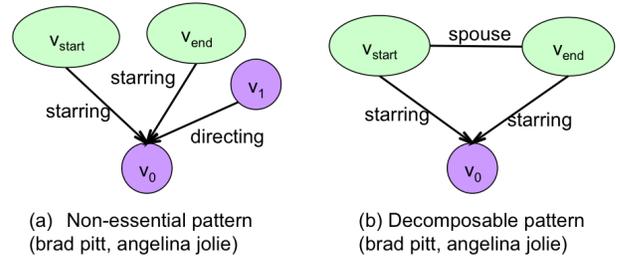

(a) Non-essential pattern
(brad pitt, angelina jolie)

(b) Decomposable pattern
(brad pitt, angelina jolie)

**Figure 5: Example non-minimal explanation patterns.**

should be integral to the connection between the target nodes. In the definition below, we give a syntactic characterization based on the graph structure of the explanation pattern.

DEFINITION 3 (ESSENTIALITY). *A node $v$ (or an edge $e$) in an explanation pattern $p = (V, E, \lambda, v_{start}, v_{end})$ is essential if there is a simple path (i.e., without repeating nodes or edges, and considering edges as undirected) through $v$ (or $e$) from $v_{start}$ to $v_{end}$. $p$ is said to be* essential *if all of its nodes and edges are essential.*

EXAMPLE 2. *Figure 5(a) shows a structure that is not essential: the node $v_1$ and the edge $(v_1, v_0)$ are not essential since they are not on any simple path from $v_{start}$ to $v_{end}$.*

Non-essential nodes and edges can be meaningful. For example, in Figure 5(a), $v_1$ provides information about the director for the movie node $v_0$, which can be interesting to users. In essence, this is akin to putting attribute constraints on the essential nodes. However, the space of non-essential graphs is extremely huge since they can be arbitrary graphs. As a result, in this paper, we will only consider explanation patterns that are essential. Non-essential nodes and edges as well as attribute constraints on essential nodes can be added in a separate stage when a candidate set of most interesting essential patterns are generated, and the details of this extension are beyond the scope of the current study.

**Non-decomposability**
The next desideratum is that we should not be able to "decompose" an explanation pattern into an equivalent set of smaller explanation patterns. From an intuitive semantic perspective, given an explanation pattern $p = (V, E, \lambda, v_{start}, v_{end})$, $p$ is decomposable if there exist two explanation patterns, $p_1 = (V_1, E_1, \lambda_1, v1_{start}, v1_{end})$ and $p_2 = (V_2, E_2, \lambda_2, v2_{start}, v2_{end})$, such that $V_1, V_2 \subset V$, and for all knowledge base instances and entity pairs, we have: $(I_{p_1} \neq \emptyset \land I_{p_2} \neq \emptyset) \Rightarrow I_p \neq \emptyset$. In another word, whenever the "sub-patterns" have some instances, then the entire pattern also must have an instance for decomposable patterns. The following is a formal definition that syntactically characterizes decomposability using the graph structure of explanation patterns.

DEFINITION 4 (DECOMPOSABILITY). *An explanation pattern $p = (V, E, \lambda, v_{start}, v_{end})$ is decomposable if there exists a partition of $E$ into $E_1, E_2$ such that $\nexists v \in V - \{v_{start}, v_{end}\}$ such that $v$ is an endpoint of an edge $e_1 \in E_1$ as well as an endpoint of an edge $e_2 \in E_2$. $p$ is said to be* non-decomposable *if it is not decomposable.*

EXAMPLE 3. *The explanation pattern in Figure 5(b) can be decomposed into two disjoint explanation patterns 4(a) and 4(b). The edge partitions of $\{(v_{start}, spouse, v_{end})\}$ and $\{(v_{start}, starring, v_0), (v_{end}, starring, v_0)\}$ do not share any nodes (besides the two target nodes).*

We combine the properties of essentiality and decomposability to denote the notion of **minimality**: An explanation pattern is said to be minimal if it is essential and non-decomposable. An explanation is said to be minimal if its explanation pattern is minimal.

## 3. EXPLANATION ENUMERATION

In this section, we study how to efficiently enumerate minimal explanations upto a limited size $n$ (provided as a system parameter) for a given node pair $v_{start}$ and $v_{end}$ in the knowledge base $G$.

One naive approach is to take advantage of existing graph enumeration algorithms [26] to generate all graph patterns and filter out the patterns that are either non-minimal or with no instances. We call this naive algorithm *NaiveEnum*, which is illustrated in Algorithm 1, and use it as the baseline in our experiments. During the enumeration, any pattern that is either duplicated (i.e., isomorphism [11] to a pattern discovered earlier) or with no instance will be pruned immediately. If the pattern is minimal, then we add it (and its instances) to the result explanation queue $Q$. However, minimality is not a pruning condition in *NaiveEnum* since non-minimal graph patterns could later be expanded to minimal graph patterns under the graph expansion rule of [26]. Not surprisingly, *NaiveEnum* is inefficient since it generates a lot of non-minimal explanation patterns and requires explicit minimality check.

---

**Algorithm 1** NaiveEnum($G, v_{start}, v_{end}, n$):$Q$

1: $Q = \emptyset, Q_p = \emptyset$
2: Append a seed pattern (a graph with a single start node) to $Q_p$
3: $i = 0$
4: **while** $i <$ length of $Q_p$ **do**
5:    $Q'_p = expand(Q_p[i])$ (Following the graph expansion rules in the graph enumeration algorithm gSpan[26], and recording the start and end node)
6:    **for** $p \in Q'_p$ **do**
7:       $I_p =$ instances of $p$ in $G$ with respect to $v_{start}$ and $v_{end}$ (can be computed efficiently from $Q_p[i]$'s instances and $G$)
8:       **if** $p$ is not duplicated $\cap |I_p| > 0 \cap |p.V| \le n$ **then**
9:          Append $p$ to $Q_p$
10:          **if** $p$ is minimal **then**
11:             Append the explanation $re = (p, I_p)$ to $Q$
12:          **end if**
13:       **end if**
14:    **end for**
15:    $i = i + 1$
16: **end while**
17: **return** $Q$

---

### 3.1 Explanation Enumeration Framework

Our goal is to design explanation enumeration algorithms that directly generate *all and only minimal explanations with at least one instance* in the knowledge base. The intuition of our algorithm comes from the observation that any minimal explanation pattern is covered by a set of path patterns, which is enforced by the essentiality property in Section 2.3, stating that each node and edge in a minimal explanation pattern must be on a single path between two target nodes. We call the set of path patterns that cover a minimal explanation pattern the *covering path pattern set* of the explanation pattern:

DEFINITION 5 (COVERING PATH PATTERN SET). *Given a minimal explanation pattern* $p_0 = (V, E, \lambda, v_{start}, v_{end})$, *we say that a multiset of path patterns* $S = \{p_1, p_2, ..., p_m\}$ *is a* covering path pattern set *if the set of path patterns in $S$ cover all the edges and nodes in $p_0$; i.e., (1) each $p_i$ ($1 \le i \le m$) maps to a simple path between $v_{start}$ and $v_{end}$ through edges in $E$, and (2) every node in $V$ and every edge in $E$ appears in at least one $p_i$ ($1 \le i \le m$).*

THEOREM 1. *Each minimal explanation pattern must have at least one covering path pattern set.*

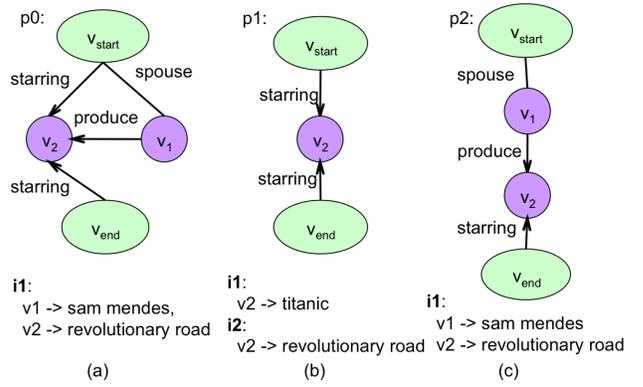

Figure 6: Example Minimal Explanations for Kate Winslet and Leonardo Dicarprio

Proofs for the theorems are omitted due to space constraints. Some minimal explanation patterns might have multiple covering path pattern sets. We also observe that we can compute the instances of a minimal explanation pattern from the instances of the path patterns in its covering path pattern set, instead of evaluating against the knowledge from scratch.

EXAMPLE 4. *The minimal explanation pattern $p_0$ in Figure 6(a) has a covering path pattern set containing the path patterns $p_1$ in Figure 6(b) and $p_2$ in Figure 6(c). Similarly, the instance $i_1$ of $p_0$ can be computed from the instance $i_2$ of $p_1$ and the instance $i_1$ of $p_2$.*

Theorem 1 suggests a general framework for minimal explanation enumeration: (1) Enumerate all path explanation patterns, including their associated instances; (2) Generate all the minimal explanation patterns (and their instances) by combining the path explanation patterns (and their instances). We only need to do explicit instance evaluation for the path explanations since instances of all other minimal explanations can be computed from them. When a pattern size limit $n$ (i.e., the number of nodes in the pattern) for a minimal explanation pattern is specified, we can derive a corresponding path pattern length limit $l$ for the covering path patterns as $l = n - 1$.

---

**Algorithm 2** GeneralEnumFramework($G, v_{start}, v_{end}, n$):$Q$

1: $Q_{path} = PathEnum(G, v_{start}, v_{end}, n - 1)$
2: $Q = PathUnion(Q_{path}, n)$
3: **return** $Q$

---

The general enumeration framework is shown in Algorithm 2. It takes $G$, $v_{start}$, $v_{end}$ and a pattern size limit $n$ as input, and returns all minimal explanation with size up to $n$. In particular, $pathEnum$ enumerates over simple path explanations (including the patterns and associated instances) for $v_{start}$ and $v_{end}$ (Section 3.2), with path pattern length up to $n-1$; all path instances are directly extracted from the knowledge base $G$. $pathUnion$ combines those simple path explanations into the minimal explanations (Section 3.3).

### 3.2 Path Explanation Enumeration

Path explanation enumeration takes $v_{start}$ and $v_{end}$ as input, a length limit $l$ and the knowledge base $G$ as parameters, and returns $Q_{path}$—the set of all path patterns for $v_{start}$ and $v_{end}$ with lengths up to $l$ (and their instances). Since path explanation enumeration can be viewed as a special case of keyword search in databases [1, 2, 3, 5, 17, 12, 13, 14, 21, 24, 29, 15] when the queried keywords match exactly two tuples, we adapt our algorithms from existing solutions instead of inventing new algorithms. There are two typical paradigms in performing keyword search in databases: (1) viewing databases as a tuple graph (tuples and their attribute values are considered as nodes and key/foreign key relationships are considered as edges) and directly searching for the instance level connecting structures[5, 17, 3, 12]; (2) first enumerating the schema level connecting structure (usually called candidate networks, akin to our path pattens here) and then evaluating the candidate networks to find out all the instances [1, 2, 13, 14, 21, 24, 29, 15]. We describe our algorithms of path enumeration following the first paradigm, since the knowledge base is already represented as a graph. Once all path instances are generated, we group them into path patterns by simply changing the nodes in the path instances to variables, a relatively straightforward process. However, algorithms and intuitions from both lines of work can be adapted into our framework.

The first path enumeration algorithm *PathEnumBasic* is adapted from *BANKS* [5]. *BANKS* runs concurrent single source shortest path algorithms from each *source* node and finds the root node connecting a set of source nodes that describe all the keywords. We apply a similar strategy to generate partial paths from both target nodes $v_{start}$ and $v_{end}$ concurrently. We generate all the path instances limited by length $\lceil l/2 \rceil$ starting from $v_{start}$ and all the path instances limited by length $\lfloor l/2 \rfloor$ starting from $v_{end}$, with shorter paths being generated first. Two path instances $i_1$ and $i_2$ from opposite directions can be connected to generate a full path instance if they end at a common node. Although this algorithm is adapted from *BANKS*, the same intuition also comes from *Discover*[14] if we are considering pattern level path enumeration: in the candidate network evaluation step of *Discover*, the optimizer iteratively chooses the most frequent (shared by most other candidate networks) "small" (number of instances is restricted by the input keywords) relations to evaluate. In our setting, this is equivalent to iteratively evaluate the shortest unevaluated path patterns connecting to any target node.

The second path enumeration algorithm *PathEnumPrioritized* is again a direct adaption from *BANKS2* [17], an improved version of *BANKS*. When generating paths from both target nodes, instead of always expanding the shortest partial paths, an activation score is used to prioritize the expanding. The activation score captures the following intuition: if expansion from one target node reaches a node with large degree, it might be very expensive to do further expansion; instead, waiting for the expansion from the other target node might be less expensive. The activation score is defined as follows: Initially, the activation score of each target node is set to 1 divided by its degree. Each time the algorithm picks a node with largest activation score to expand the paths ending at that node. During the expansion, activation score of the node spread to its none-target node neighbors (the activation score spread to each new node is set to the activation score of the original node divided by the degree of new node) and the activation score of original node is set to 0. For each none-target node, activation scores provided by different neighbors are added up. If a node receives activation scores from both target nodes, it indicates the identification of new connecting paths. Again, if our algorithm was adapted from candidate network generation and enumeration, the intuition for the same strategies comes from *Discover*[14] when we assume $b > 0$ in the cost model (i.e., we take into consideration the estimated size of join results) for candidate network evaluation.

## 3.3 Path Explanation Combination

Path explanation combination takes the length-limited path explanations $Q_{path}$ as input, the pattern size limit $n$ as parameter, and return $Q$—the set of all minimal explanations with limited pattern size. Combining path explanations to generate minimal explanations is a non-trivial task. Any set of path explanation patterns could be a covering path pattern set for some minimal explanation patterns and there are many ways of combining path patterns in a covering path pattern set. In order to have a better understanding of how we can generate all the minimal explanation patterns (and hence the explanations), we partition the set of all minimal explanation patterns $MinP$ into disjoint sets, depending on the minimal cardinality (number of path patterns) of any covering path pattern set of a minimal explanation pattern:

$$MinP = \{MinP(k), k = 1..\infty\}, \quad (1)$$

where $MinP(k)$ represents the set of minimal explanation patterns with minimal covering path pattern set cardinality of $k$. In particular, $MinP(1)$ represents all path patterns. We can extend the notion of covering path pattern set to include non-path minimal patterns:

DEFINITION 6 (COVERING PATTERN SET). *Given a minimal explanation pattern $p_0 = (V, E, \lambda, v_{start}, v_{end})$, we say that a multiset of patterns $S = \{p_1, p_2, ..., p_m\}$ is a* covering pattern set *if the set of patterns in S cover all the edges and nodes in $p_0$; i.e., (1) each $p_i$ ($1 \leq i \leq m$) maps to a sub-component of $p_0$ connecting $v_{start}$ and $v_{end}$ through edges in E, and (2) every node in V and every edge in E appears in at least one $p_i$ ($1 \leq i \leq m$).*

Just like covering path pattern set, given a knowledge base, the instances of a minimal pattern can be computed from instances of patterns in its cover pattern set. The following theorem shows that $MinP(k), k > 1$ can be derived from minimal patterns with smaller cardinality:

THEOREM 2. *Each explanation pattern in $MinP(k)$ ($k > 1$) must have a covering pattern set composed of a pattern in $MinP(k-1)$ and a pattern in $MinP(1)$.*

Theorem 2 suggests that starting from $MinP(1)$, we could iteratively enumerate $MinP(k), k > 1$ from $MinP(k-1)$ and $MinP(1)$. Our first path explanation combination algorithm *PathUnionBasic* (Section 3.3.1) directly applies this finding to reduce the enumeration space. In Section 3.3.2 we discuss additional pruning opportunities for *PathUnionBasic* and propose an even more efficient combination algorithm *PathUnionPrune*.

### 3.3.1 PathUnionBasic

Algorithm 3 illustrates the pseudocode for *PathUnionBasic*, and we explain its critical components as follows:

**Enumeration (Line 1 - Line 15):** Path explanations in $Q_{path}$ are used as the seed explanations and put in an explanation queue $Q$. For each explanation $re$ in $Q$, the algorithm combines $re$ with each path explanation in $Q_{path}$ to generate new minimal explanations. The *Explanation Merging* component ensures that the generated explanation patterns are minimal and each is associated with at least 1 instance. The *Duplication Checking* component ensures that only unique explanations are appended to $Q$ (i.e., duplicates are pruned). The process stops when all explanations in $Q$ have been expanded and no more explanations can be generated. All the minimal explanations with limited pattern size are guaranteed to be in $Q$ at the end of the process. (Proof omitted due to space constraints.)

**Explanation Merging (Line 24 - Line 41):** To define the merge of two explanations, we consider a partial one-to-one mapping between the patterns of two explanations, say $p_1 = (V_1, E_1, \lambda_1, v1_{start}, v1_{end})$ and $p_2 = (V_2, E_2, \lambda_2, v2_{start}, v2_{end})$:

**Algorithm 3** PathUnionBasic($Q_{path}$,n):Q
―――――――――――――――――――――――――――――――――――
 1: $Q = Q_{path}$; $Q_{expand} = Q_{path}$
 2: **while** $Q_{expand} \neq \emptyset$ **do**
 3:     $Q_{new} = \emptyset$
 4:     **for all** $(re_1, re_2)$ pair *in* $Q_{expand} \times Q_{path}$ **do**
 5:         $Q_{temp} = merge(re_1, re_2, n)$
 6:         **for** $re \in Q_{temp}$ **do**
 7:             **if** $duplicated(re, Q \cup Q_{new}) = False$ **then**
 8:                 Append $re$ to $Q_{new}$
 9:             **end if**
10:         **end for**
11:     **end for**
12:     Append $Q_{new}$ to $Q$
13:     $Q_{expand} = Q_{new}$
14: **end while**
15: **return** $Q$

16: **function** duplicated($re,Q$):$duplicated$
17:     **for** $re_1 \in Q$ **do**
18:         **if** exist an ismorphism between $re$'s pattern and $re_1$'s pattenrn **then**
19:             **return** $True$
20:         **end if**
21:     **end for**
22:     **return** $False$
23: **end function**

24: **function** merge($re_1,re_2,n$):$Q_{new}$
25:     $(p_1, I_{p_1})$ = $re_1$'s pattern and instances
26:     $(p_2, I_{p_2})$ = $re_2$'s pattern and instances
27:     $Q_{new} = \emptyset$
28:     **for all** partial one-to-one mapping $f$ from $p_1.V$ to $p_2.V$ **do**
29:         $p_{new} = p_1 \cup_f p_2$
30:         $I_{p_{new}} = \emptyset$
31:         **for all** $(i_1, i_2)$ pair *in* $I_{p_1} \times I_{p_2}$ **do**
32:             **if** $i_1, i_2$ is the same on every pair of matched nodes **then**
33:                 Append $i_{new} = i_1 \cup_f i_2$ to $I_{p_{new}}$
34:             **end if**
35:         **end for**
36:         **if** $|p_{new}.V| \leq n$ and $|I_{p_{new}}| > 0$ **then**
37:             Append $re_{new} = (p_{new}, I_{p_{new}})$ to $Q_{new}$
38:         **end if**
39:     **end for**
40:     **return** $Q_{new}$
41: **end function**

(1) $v1_{start}$ and $v1_{end}$ should be mapped to $v2_{start}$ and $v2_{end}$ respectively.

(2) A non-target node $v_1 \in V_1 - \{v1_{start}, v1_{end}\}$ of $p_1$ could be mapped to a non-target node $v_2 \in V_2 - \{v2_{start}, v2_{end}\}$ of $p_2$ or does not map to any node. (Same restriction for $v_2$)

(3) One-to-one mapping is enforced (when there is a mapping).

(4) At least one non-target node of $p_1$ should be mapped to a non-target node of $p_2$.

Given this partial one-to-one mapping function $f$, a new explanation pattern can be merged from $p_1$ and $p_2$ following the mapping function $f$. We use an operator $\cup_f$ to represent this merging: nodes and edges in both patterns should be put into the new pattern, with each pair of matched nodes merged as one node. If there are multiple edges with same label between a pair of nodes in the new pattern, they are merged as well. Since each node and edge of the new pattern are coming from two minimal explanation patterns, it is guaranteed to be on a single path between target nodes. Therefore the new pattern is essential. On the other hand, requirement (4) of the mapping guarantees that the new pattern is also non-decomposable. Therefore the new explanation pattern is minimal. The instances of the new explanation can be generated by enforcing the same mapping on each pair of instances from $re_1$ and $re_2$, with the requirement that two instances agree on every pair of matched nodes. The new explanation is kept only if it has at least one instance.

EXAMPLE 5. *Consider the two patterns $p_1$ in Figure 6(b) and $p_2$ in Figure 6(c). A valid partial one-to-one mapping between the two patterns is $p_1.v(start)$—$p_2.v(start)$, $p_1.v(end)$—$p_2.v(end)$, nothing—$p_2.v1$, $p_1.v2$—$p_2.v2$. Combining $p_1$ and $p_2$ following the mapping yields the pattern $p_0$. $p_0$'s instance $i_1$ can be computed from $i_2$ of $p_1$ and $i_1$ of $p_2$ following the mapping.*

**Duplication Checking (Line 16 - 23):** An explanation could be generated multiple times during the enumeration (e.g., combination of different pairs of minimal explanations could yield the same minimal explanation). We perform duplication check for a new explanation by checking graph isomorphism [11] of its explanation pattern against patterns of any existing explanations. If a graph isomorphism is detected, then the new explanation is duplicated and therefore ignored.

### 3.3.2 PathUnion with Pruning

*PathUnionBasic* generates all but only the minimal explanations with at least 1 instance. Therefore it is much more efficient than the baseline algorithm. However, since a minimal explanation might be generated multiple times during the enumeration (indicating some of the combinations might be unnecessary), the efficiency of the algorithm is still restricted by the number of times we need to merge the minimal explanations. The following theorem allows us to decrease the number of merges required:

THEOREM 3. *Each explanation pattern in $MinP(k), (k > 2)$ must have a covering pattern set $\{p_1, p_0\}$ of size 2, such that $p_0, p_1 \in MinP(k-1)$, and $p_0$ and $p_1$ share a $MinP(k-2)$ subcomponent $p_2$. i.e., the pattern graph of $p_2$ is a subgraph of patterns of $p_0$ and $p_1$, and start and end node of $p_2$ map to start and end node of $p_0$ and $p_1$.*

Another way to interpret this theorem is that: Let $p_1 \in MinP(k)$ ($k > 2$), $p_2 \in MinP(k-1)$ and $p_5 \in MinP(1)$. In order to generate $p_1$ from $p_2$ and $p_5$, there must be $p_3$ and $p_4$ that satisfy following conditions: $p_3 \in MinP(k-1)$; $p_4 \in MinP(k-2)$ and is a subcomponent of $p_2$; $p_3$ can be merged from $p_4$ and $p_5$. Based on this interpretation, we can reduce the number of times we need to combine a minimal explanation with a path explanation. Specifically, during the enumeration, for each explanation in $Q$ that has its pattern $p_2$ in in $MinP(k-1)$, we record the pairs of $p_4 \in MinP(k-2)$ and $p_5 \in MinP(1)$ (and hence the corresponding explanations) it was generated from. For an explanation with its pattern $p_2 \in MinP(k-1)$, by comparing the composition history with other explanations that have patterns in $MinP(k-1)$ and enforcing the requirement from Theorem 3, we can decide whether the subset of paths should be merged with $p_2$. The pseudocode of the enumeration algorithm with pruning is in Algorithm 4 and we call this algorithm *PathUnionPrune*. We use queues $H_{expand}$ and $H_{new}$ to store the composition history for $MinP(k-1)$ and $MinP(k)$'s corresponding explanations respectively.

## 4. INTERESTINGNESS MEASURES AND EXPLANATION RANKING

When the number of minimal explanations is larger than what we can expect users to consume, it is important to *rank* them in order of their "interestingness." This interestingness measure can be defined in a variety of different ways and is often subjective. In this paper, we aim to present a comprehensive set of such measures and design

246

**Algorithm 4** PathUnionPrune($Q_{path}$,n):Q

1: $Q = Q_{path}$; $Q_{expand} = Q_{path}$
2: **while** $Q_{expand} \neq \emptyset$ **do**
3:    $Q_{new} = \emptyset$; $H_{new} = \emptyset$
4:    **for all** $i_1$ in [0 .. $length(Q_{expand})$ - 1] **do**
5:       $S_{path} = \emptyset$
6:       **if** $Q_{expand} = Q_{path}$ **then**
7:          $S_{path}$ = [0 .. $length(Q_{path})$ - 1]
8:       **else**
9:          **for all** $i_2$ in [0 .. $length(Q_{expand})$ - 1] **do**
10:             **for all** $((x, j_1),(x, j_2))$ pair in $H_{expand}[i_1] \times H_{expand}[i_2]$ **do**
11:                Add $j_2$ to $S_{path}$
12:             **end for**
13:          **end for**
14:       **end if**
15:       **for all** $i_2$ in $S_{path}$ **do**
16:          $Q_{temp} = merge(Q_{expand}[i_1], Q_{path}[i_2], n)$
17:          **for** $re \in Q_{temp}$ **do**
18:             **if** $duplicated(re, Q) = False$ **then**
19:                **if** $duplicated(re, Q_{new}) = False$ **then**
20:                   Append $re$ to $Q_{new}$
21:                   Append $\emptyset$ to $H_{new}$
22:                **end if**
23:                $i_{re}$ = $re$'s index in $Q_{new}$
24:                Append $(i_1, i_2)$ to $H_{new}[i_{re}]$
25:             **end if**
26:          **end for**
27:       **end for**
28:    **end for**
29:    Append $Q_{new}$ to $Q$
30:    $Q_{expand} = Q_{new}$; $H_{expand} = H_{new}$
31: **end while**
32: **return** $Q$

efficient algorithms for computing them. In Section 5, we conduct user studies to analyze the effectiveness of our proposed measures.

We start by formally defining a generic interestingness measure. We pay particular attention to one of the key properties of a measure, namely *monotonicity*. We shall see that anti-monotonicity, which holds for some of our measures, can be used for pruning in enumeration and ranking of explanations.

DEFINITION 7 (MEASURE AND MONOTONICITY).
*An interestingness measure $\mathcal{M}$ is a function that takes as input the knowledge base $G = (V, E, \lambda)$, an explanation pattern $p = (V', E', \lambda', v'_{start}, v'_{end})$, and target nodes $v_{start}, v_{end} \in G.V$ and returns a number $\mathcal{M}(G, p, v_{start}, v_{end}) \in \mathbb{R}$.*

*We say that a measure $\mathcal{M}$ is monotonic (anti-monotonic, resp.) if and only if $\mathcal{M}(G, p_1 = (V'_1, E'_1, \lambda'_1, v'_{start}, v'_{end}), v_{start}, v_{end}) \geq (\leq, resp.) \mathcal{M}(G, p_2 = (V'_2, E'_2, \lambda'_2, v'_{start}, v'_{end}), v_{start}, v_{end})$ whenever the graph $G_2$ induced by $V'_2, E'_2, \lambda'_2$ is a subgraph of $G_1$ induced by $V'_1, E'_1, \lambda'_1$.*

Note that although an interestingness measure is defined in terms of an explanation pattern, by including the knowledge base as one of the inputs to the measure function, the corresponding instances can also be derived. Therefore, an interestingness measure actually measures the interestingness of explanations.

Most existing measures for connecting structures is derived from their topological structures; examples of them include the size measure and random walk measure, which we will discuss in Section 4.1. However, these measures do not capture the aggregated information of the instances, e.g., co-starred in 10 movies. Therefore, we propose two novel families of interestingness measures: *aggregate* measures and *distributional* measures. Aggregate measures are obtained by aggregating over individual instances. One intuitive aggregate measure is the *count* measure, where the interestingness of an explanation is proportional to the number of explanation instances obtained by applying the explanation pattern to the knowledge base. We can compare simple aggregate measures against those of other pairs of entities to produce *distributional* measures. We describe aggregate and distributional measures in Sections 4.2 and 4.3 respectively.

## 4.1 Structure-based measures

The structure of an explanation pattern can affect the interestingness of an explanation. These kinds of interestingness measures are frequently used in existing works [1, 2, 3, 5, 17, 12, 13, 14, 21, 15, 8, 10, 18, 22, 25]. We describe two representatives in this section: the size measure and the random walk measure. Size of pattern is a simple but useful summarization of the structural interestingness, and it can be easily used together with any other interestingness measure. Another structural interestingness measure we consider is based on an extension of the random walk process described in [10]: each connecting instance graph is regarded as an electrical network (e.g. each edge represents a resistor) and the amount of current delivered from the start entity to the end entity is used as the interestingness of the connecting graph. In our case, we apply the random walk on the pattern and use the result as the interestingness measure for the explanation.

## 4.2 Aggregate Measures

Aggregate measures follow the intuition that the more instances an explanation has, the more interesting it is. For example, consider the explanation in Figure 4(b) (co-starring): the more movie instances $v_0$ can map to, the higher the aggregate measure is, and the more interesting the explanation is. We distinguish two ways of aggregating the number of instances: *count* and *monocount*.

*Count*

The *count* measure simply gives the total number of distinct instances an explanation has. Formally, we have:

$$\mathcal{M}_{count}(G, p, v_{start}, v_{end}) = |\{f | f \text{ satisfies Definition 2}\}|$$

While intuitive to define, $\mathcal{M}_{count}$ is neither monotonic nor anti-monotonic[4], which makes it difficult to compute due to the lack of pruning possibilities.

*Monocount*

To address the shortcoming of $\mathcal{M}_{count}$, we propose an alternative count measure that has the anti-monotonicity property. Given $G$, $p = (V', E', \lambda, v'_{start}, v'_{end})$ and the target nodes, let $uniq(v)$, $v \in V'$ denote the number of distinct assignments that can be made to any variable over all instances:

$$uniq(v) = |\{f(v) | f \text{ satisfies Definition 2}\}|$$

The monocount of $p$ gives the fewest number of assignments over all variables (except the two target nodes):

$$\mathcal{M}_{monocount}(G, p, v_{start}, v_{end}) = \min_{v \in p.V' - \{v'_{start}, v'_{end}\}} uniq(v)$$

We override the above formula and define monocount to be 1 in the special case that there is a direct edge between the target entities.

EXAMPLE 6. *Let us assume that in Figure 6(a), there is another instance with $v_1$ mapping to "sam mendes" and $v_2$ mapping*

to "revolutionary road II". Then in this case $|uniq(v_1)| = 1$ and $|uniq(v_2)| = 2$, therefore $min_{v_i}(|uniq(v_i)|) = 1$ and the monocount is 1. In comparison, the count would be 2 in this case.

Note that when there is a single non-target variable, $\mathcal{M}_{monocount} = \mathcal{M}_{count}$. Our measure is an extension of the anti-monotonic support of sub-graphs within a single graph that was introduced in [6].

### 4.3 Distribution-Based Measures

Aggregate measures are suitable for comparing explanations for a given pair of target entities. However, they do not capture the "rarity" of an explanation across different pairs of target entities. For example, a spousal relationship always has a count of 1, but it is arguably more interesting than a co-starring relationship with a count of 1. This is because co-starring relationships are much more common than the spousal relationships. To capture such rarity information, we propose two distributional measures—*local* and *global*—that compare the aggregate measure of an explanation against the aggregate measures of a set of explanations obtained by varying the target nodes. [5]

Let $\mathcal{M}_{agg}$ be the specific aggregate measures we adopt, and $\{a_1, a_2, \ldots, a_n\}$ be the sequence of $\mathcal{M}_{agg}$ values in increasing order, local and global distributions $D^l = \{(a_i^l, c_i^l)\}$ and $D^g = \{(a_i^g, c_i^g)\}$ can be defined below, where the former is obtained by varying only the *end* target node and the latter is obtained by varying *both* target nodes:

$$c_i^l = |y \in G.V \mid \mathcal{M}_{agg}(G, p, v_{start}, y) = a_i^l|$$

$$c_i^g = |(x, y) \in (G.V \times G.V) \mid \mathcal{M}_{agg}(G, p, x, y) = a_i^g|$$

Intuitively, $c^l$ and $c^g$ give the number of entity pairs whose explanations produce the aggregate values of $a^l$ and $a^g$ respectively. The entire distribution of these count values is then used to compute the rarity of the given explanation and entity pair using standard statistical techniques. In particular, we compute the *position* of the given explanation with respect to the distribution: Let $A$ be the value of $\mathcal{M}_{agg}$ for the given explanation and $D = \{(a_1, c_1), \ldots, (a_n, c_n)\}$ be the distribution to be compared against, we have:

$$\mathcal{M}_{position} = \sum_{i|a_i > A} c_i$$

Another alternative is to count how many standard deviations $A$ is away from the mean of $D$, which turns out to be similarly effective as $\mathcal{M}_{position}$. We ignore the details here due to space constraints.

EXAMPLE 7. *Consider the co-starring explanation (Figure 4(b)) for Brad Pitt and Angelina Jolie. The corresponding count is 1 since they co-starred in only 1 movie. The local distribution of counts for Brad Pitt and any other actor/actress is shown as follows:*

$$D^l = \{(1, 130), (2, 8), (3, 10), (4, 2)\}$$

*Therefore the corresponding position in the local distribution is $8 + 10 + 2 = 20$. In contrast, their spousal explanation (Figure 4(a)) also has a count of $1$. However, its position in the local distribution is $0$ since no other person with Brad Pitt has a larger count for a spousal relationship. Therefore by comparing the positions in the local distribution we can infer that the spousal explanation is more interesting than the co-starring explanation.*

---

[5] Although used in a completely different domain, aggregated measures and distribution-based measures are analogous to the TF-IDF measure in IR.

### 4.4 Explanation Ranking

In this section we discuss how to efficiently rank the explanations given a pair of target entities. Specifically, given an interestingness measure and a parameter $k$, the explanation ranking algorithm returns a ranked list of top-$k$ most interesting explanations based on the interestingness measure.

---

**Algorithm 5** GeneralRankFramework($G$, $v_{start}$, $v_{end}$, $n$, $\mathcal{M}$, $k$):Q

1: $Q = GeneralEnumFramework(G, v_{start}, v_{end}, n)$
2: $Q_{int} = \emptyset$
3: **for** $re \in Q$ **do**
4:     Append $\mathcal{M}(G, re.pattern, v_{start}, v_{end})$ to $Q_{int}$
5: **end for**
6: Sort $Q$ based on $Q_{int}$
7: $Q$ = first $k$ entries in $Q$
8: return $Q$

---

Algorithm 5 illustrates the general ranking framework, which involves three steps: explanation enumeration (based on Section 3), interestingness computation, and explanation ranking. This general ranking algorithm can be applied to all interestingness measures discussed in the previous subsections.

For certain interestingness measures, however, we can design more efficient ranking algorithms: increased efficiency can be obtained by aggressively pruning explanations while interleaving the enumeration, interestingness computation, and ranking steps. The pruning for distribution based measures is described in Section 5.3.2. Here, we briefly describe the case of ranking based on anti-monotonic interestingness measures.

Recall the anti-monotonicity property from Section 4 (which *monocount* measure satisfies); the following theorem allows us to prune enumerations when considering anti-monotonic measures.

THEOREM 4. *Given the knowledge base $G = (V, E, \lambda)$ and target nodes $v_{start}, v_{end}$, and anti-monotonic interestingness measure $\mathcal{M}$, suppose a relationship explanation $re' = (p', I')$ is derived from relationship explanation $re = (p, I)$ using PathUnionBasic (Algorithm 3) or PathUnionPrune (Algorithm 4). We then have that $\mathcal{M}(G, p, v_{start}, v_{end}) \geq \mathcal{M}(G, p', v_{start}, v_{end})$. (Thereforem if $re$ is not among the top-$k$ most interesting explanations, no $re'$ derived from it is.)*

Intuitively, any expansion of an explanation can only reduce the value of an anti-monotonic measure. Using the theorem, we can integrate the three steps of the general ranking algorithm by maintaining a current top-$k$ list of most interesting explanations during enumeration. Upon generation of each explanation, we perform the following steps:

**Step 1**: Calculating the interestingness of the explanation.
**Step 2**: Updating the top-$k$ list of explanations; explanations not in the top-$k$ list are pruned out.
**Step 3**: Continue expansion only from the current set of top-$k$ explanations.

Finally, the top-$k$ most interesting explanations are returned. Intuitively, this algorithm is more efficient than the general ranking algorithm since fewer explanations are enumerated, and this intuition is supported by our experimental evaluation (Section 5).

## 5. EXPERIMENTS

We implemented the *REX* system in Python and performed extensive experiments using a real world knowledge base to evaluate its efficiency and effectiveness. Specifically, we analyze the performances of explanation enumeration algorithms and ranking algorithms in Sections 5.2 and Section 5.3, respectively. We also perform extensive quality assessments based on detailed user studies (Section 5.4) to verify the necessity of our explanation definition (e.g., including non-path explanations) and the effectiveness of explanations generated by *REX*. All experiments are performed on a MacBook Pro with 2.53 GHz Dual Core CPU and 4GB RAM.

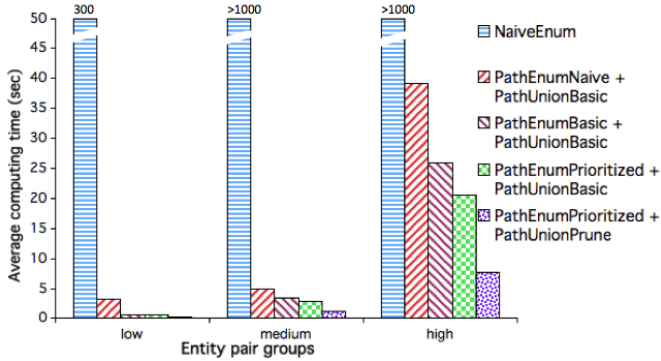

**Figure 7: Compare explanation enumeration algorithms.**

## 5.1 Experimental Settings

**Knowledge Base**: We extracted from DBpedia (http://dbpedia.org/) all entertainment related entities and relationships to form our experiment knowledge base. There are a total of 20 entity types and 2,795 primary relationship types. Overall, the knowledge base contains 200K entities and over 1.3M primary relationships.

**Target Entity Pairs**: We generate related entities for evaluation as follows: we randomly select an entity as the start entity from the knowledge base and then randomly select one of its related entities as suggested by the search engine[6]. We categorize the pairs based on their "connectedness", which is computed by the number of simple paths that connect the two entities within a given length limit[7]: *low* (connectedness: 0 - 30), *medium* (connectedness: 30 - 100), and *high* (connectedness > 100). From each of the three groups, we randomly pick 10 related pairs; these 30 related entity pairs are used for performance evaluation.

## 5.2 Performance of Enumeration Algorithms

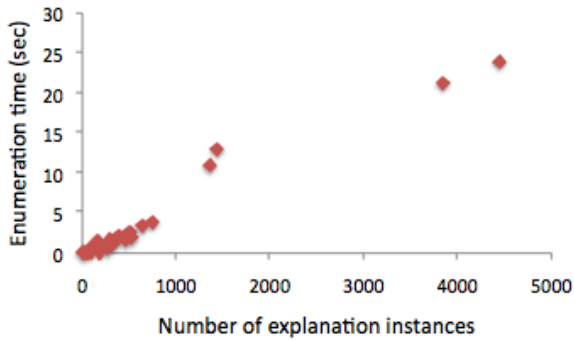

**Figure 8: Explanation enumeration time vs. number of explanation instances.**

In this section, we compare the performance of our minimal explanation enumeration algorithms. As discussed in Section 3, there are 3 types of optimizations we consider: (a) using path enumeration and union framework instead of graph enumeration, (b) picking the best path enumeration algorithm from existing solutions, (c) optimizing the path union algorithm. To illustrate the usefulness of each optimization decision, we consider the following combinations: 1. *NaiveEnum* (using graph enumeration, note that graph enumeration cannot be used in combination with the other two types of optimizations), 2. *PathEnumNaive*[8] + *PathUnionBasic* 3. *PathEnumBasic* + *PathUnionBasic* (using path enumeration and union framework with baseline algorithms for both components), 4. *PathEnumPrioritized* + *PathUnionBasic* (using prioritized path enumeration algorithm with basic path union algorithm), 5. *PathEnumPrioritized* + *PathUnionPrune* (using improved path enumeration and union algorithms). We set the pattern size limit to 5 in the experiments.

---

[6] http://search.yahoo.com/
[7] We set the length limit to 4 to match the pattern size limit of 5 in our experiments.

Figure 7 shows the efficiencies of different explanation enumeration algorithms. Any combination of the path enumeration and union algorithm, including the most naive version *PathEnumNaive* + *PathUnionBasic*, shows orders of magnitude improvement over *NaiveEnum*, for all three entity pair groups (low, medium and high). This demonstrates the efficiency of our framework, which does not generate any non-minimal structure during the enumeration. The comparison of *PathEnumBasic* + *PathUnionBasic* and *PathEnumPrioritized* + *PathUnionBasic* indicates *PathEnumPrioritized* is slightly more efficient than *PathEnumBasic*. (And both of them are better than *PathEnumNaive* as expected.) Although this improvement is not our contribution, the result tells us which is the best path enumeration algorithm to choose. Finally, the comparison of *PathEnumPrioritized* + *PathUnionBasic* and *PathEnumPrioritized* + *PathUnionPrune* shows that *PathUnionPrune* is more efficient than *PathUnionBasic* due to the additional shared-component pruning performed during the enumeration process: on average, by using *PathUnionPrune*, it takes only one third of the time of when using *PathUnionBasic*.

Figure 8 shows the enumeration time (using algorithm *PathEnumPrioritized* + *PathUnionPrune*) for all 30 entity pairs, where x-axis is the number of explanation instances for the pair and y-axis is the enumeration time. The enumeration time increases linearly with the number of explanation instances between the pairs, which reaches as high as 5000, demonstrating the scalability of the *REX* system[9].

## 5.3 Performance of Ranking Algorithms

In this section we evaluate the performance of ranking algorithms. The running time with ranking is affected by two components: the time for enumeration and the time for computing the measure. For simple aggregate measures such as count and monocount, the enumeration time dominates. However, for distributional measures, measure computation takes longer (because the same measure needs to be computed for additional sample entity pairs). We show that our pruning algorithms successfully improve the performances for all measures, either through reducing enumeration time or measure computation time.

### 5.3.1 Top-k Pruning for Anti-monotonic Measures

Figure 9 shows the effects of top-$k$ ($k = 10$) pruning for the measure $\mathcal{M}_{monocount}$, following the top-k pruning algorithm for anti-monotonic measures discussed in Section 4.4. In all cases, top-$k$ pruning reduces the running time to under 0.5 seconds, and it is sometimes several hundred times more efficient than full enumeration. In Figure 10, we examine how different values of $k$ affect the running time. As expected, when $k$ is very small, using top-k pruning significantly improves efficiency. As $k$ becomes larger, the improvement diminishes. When $k$ is very large, the pruning algorithm is close to (and in the medium group slower than) the non-pruning algorithm, since very few results are pruned and maintaining the top-$k$ list adds overhead.

---

[8] *PathEnumNaive* is a most naive path enumeration algorithm: it enumerates all length-limited paths from start entity and checks if each path ends at the end entity. It is worse than any existing solution therefore we do not include it in Section 3.2 as the baseline. However, because it uses the most naive design without any optimization, its improvement over *NaiveEnum* shows the benefits of adopting our framework.

[9] It is worth noting that density rather than the total size of the knowledge base affects the performance of enumeration. Therefore the performance would not be affected much even if we adopt the full DBPedia knowledge base in our experiments.

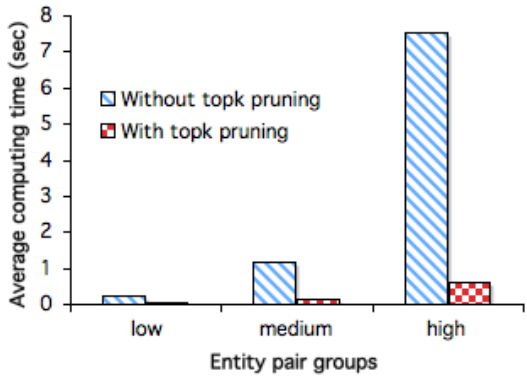

**Figure 9: Effect of top-k (k = 10) pruning on monocount computing**

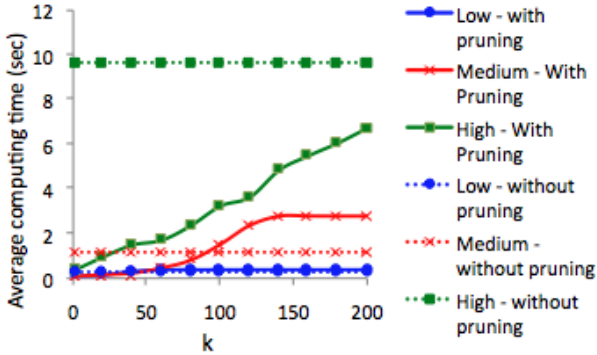

**Figure 10: Average compute time for different k in top-k pruning**

### 5.3.2 Computing and Pruning for Distribution-Based Measures

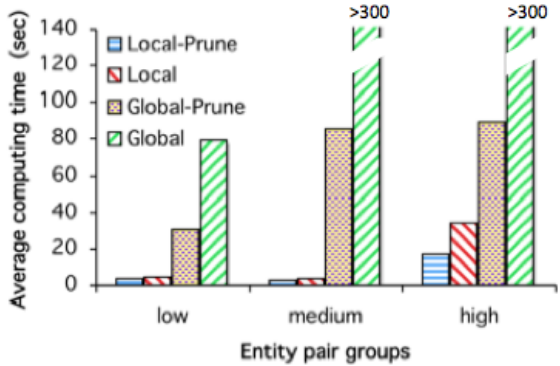

**Figure 11: Average time for computing top-10 explanation using distribution-based measure $\mathcal{M}_{position}$.**

Despite the fact that distribution-based measures as described in Section 4.3 are not anti-monotonic and therefore not subject to the aggressive pruning introduced in Section 4.4, we can potentially optimize their computation by integrating the measure computation with explanation ranking. Here, we use the local distribution-based position measure to illustrate how the pruning can be done.

Specifically, given a pair of target nodes $v_{start}$ and $v_{end}$, the knowledge base $G$ with all the primary relationships stored in a relational table $R(eid1, eid2, rel)$[10], an explanation pattern $re$, and its $\mathcal{M}_{count}$ $c$, the local distributional position of $re$ based on $\mathcal{M}_{count}$ can be computed via evaluating a SQL query describing $re$'s pattern. Assuming $re$ is the *co-starring* relationship, the corresponding SQL statement is as follows:

```
SELECT v_start, R2.eid1, count(*) as count
FROM R as R1, R as R2
WHERE v_start = R1.eid1 AND R1.eid2 = R2.eid2
      AND R1.rel = 'starring'
      AND R2.rel = 'starring'
GROUP BY v_start, R2.eid1
HAVING count > c
```

The structure of the explanation pattern is encoded in the "FROM" and "WHERE" clauses (e.g., each edge would be mapped to a table in the "FROM" clause). Each returned record represents a pair of entities (within the local distribution) that have count greater than the target entity pair. Therefore, the number of records in the SQL statement gives the desired position of the explanation.

To improve upon the general brute force Algorithm 5, we maintain a top-$k$ list of explanations when computing the interestingness of the explanations and modify the SQL query above for optimization. For example, if we know the current $k^{th}$ most interesting explanation has a position of $p$, then we needn't compute the position for target entities whose position is guaranteed to be above $p$. This optimization can be reflected by simply adding a *LIMIT p* clause in the SQL query above.

We implemented this pruning strategy and evaluated its effectiveness for top-$k$ ($k = 10$) explanation ranking using distribution-based measure $\mathcal{M}_{position}$. There are four different scenarios: local distribution, local distribution with pruning, global distribution, and global distribution with pruning. Since the true global distribution would be prohibitively time-consuming to compute, we use 100 local distributions to estimate the global distribution, with each local distribution associated with randomly chosen start entities. The computation time in all four scenarios are shown in Figure 11. First, we note that pruning is beneficial regardless whether the measure is local or global distribution based. In particular, pruning can speed up the computation by 2 times for local distributional measures. However, ranking using global distributional measure is still quite costly even with pruning. We note that the cost of computing distributional measures can be further decreased by amortizing the computation over different pairs by sharing the computation involved. Also, distributional measures can be computed in parallel as count for different node pairs can be computed separately. Finally, combination of distributional measures with other measures could decrease the computation time. For example, we can use some other measure (e.g., size) as the primary comparison index and use distributional measures only to tie-break the less expensive primary index comparison. Our experiments show that in average computation time based on such combinational measures are several times faster than using distributional measures alone.

## 5.4 Measure Effectiveness

In this section, we analyze the effectiveness of explanations generated by *REX*. In Section 5.4.1, we compare the relative effectiveness of different interesting measures and their combinations.

---

[10]The knowledge base can be stored using other data models (e.g, $RDF$), and the same computing strategy can still be applied.

| Measure | P1 | P2 | P3 | P4 | P5 | Avg |
|---|---|---|---|---|---|---|
| size | 50 | 51 | 33 | 51 | 52 | 47 |
| random-walk | 55 | 45 | 41 | 45 | 47 | 47 |
| count | 53 | 39 | 38 | 53 | 45 | 46 |
| monocount | 54 | 40 | 40 | 52 | 41 | 45 |
| local-dist | 62 | 47 | 53 | 58 | 59 | 55 |
| global-dist | 61 | 37 | 58 | 61 | 58 | 55 |
| size + monocount | 67 | 60 | 50 | 61 | 59 | 59 |
| size + local-dist | 67 | 60 | 50 | 62 | 60 | 60 |

**Table 1: Comparing different interestingness measures.**

In Section 5.4.2 we show why only using path is not sufficient to model all possible interesting explanations.

### 5.4.1 Effectiveness of Interestingness Measures

We compare the 6 measures discussed in Section 4: size ($\mathcal{M}_{size}$), random walk ($\mathcal{M}_{walk}$), count ($\mathcal{M}_{count}$), monocount ($\mathcal{M}_{monocount}$), position in local and global distributions ($\mathcal{M}_{position}^{local}$, $\mathcal{M}_{position}^{global}$). We also expect that combinations of different measures, especially combinations of structure based measures (e.g., $\mathcal{M}_{size}$) with aggregated and distributional measures (e.g., $\mathcal{M}_{count}$, $\mathcal{M}_{monocount}$, $\mathcal{M}_{position}^{local}$, $\mathcal{M}_{position}^{global}$), could be very helpful since they try to capture the interestingness of explanations from different while complementary directions. Therefore, we also include some combinational measures in the result to verify the idea.

We randomly selected 5 entity pairs for this study [11]: *P1: (brad pitt, angelina jolie), P2: (kate winslet, leonardo dicaprio), P3: (tom cruise, will smith), P4: (james cameron, kate winslet), P5: (mel gibson, helen hunt)*. For each pair, each measure is used to rank the top-10 most interesting explanations. The resulting explanations are randomized and mixed together so the user can't tell how an explanation is measured by each measure. The user is then asked to label each explanation as very relevant (score 2), somewhat relevant (1), or not relevant (0). For each ranking methodology, a DCG-style score [12] is computed as follows:

$$score(M) = m\Sigma_i(w_i \times s_i), i \in [1, 10]$$

where $m$ is a normalization factor to ensure the scores fall within [0, 100], $w_i$ are the weights given for each rank position (in our case, $w_i = 1/\log_2(i+1)$)[13], and $s_i$ are the individual explanation scores at position $i$ as ranked by the corresponding measure.

A total of 10 users responded to our user study. The average scores of different measures for each entity pair are shown in first 6 lines in Table 1. The effectiveness of $\mathcal{M}_{size}$, $\mathcal{M}_{walk}$, $\mathcal{M}_{count}$ and $\mathcal{M}_{monocount}$ are very similar. (The most simple size measure is even slightly better.) As we expected, the two distribution-based measures are statistically better than the simple aggregate measures and structure based measures. It is interesting to see that, despite its much more limited sampling scope, $\mathcal{M}_{position}^{local}$ performs as well as $\mathcal{M}_{position}^{global}$ in terms of ranking quality. Given that $\mathcal{M}_{position}^{local}$ is much cheaper to compute (Figure 11), we recommend that $\mathcal{M}_{position}^{local}$ be always used in place of $\mathcal{M}_{position}^{global}$ if distribution-based measures are desired.

We also consider two very simple combinations of the measures: $\mathcal{M}_{size\&monocount}$ (using $\mathcal{M}_{size}$ as the primary comparison index and use $\mathcal{M}_{monocount}$ as the secondary comparison index), $\mathcal{M}_{size\&local-dist}$ (using $\mathcal{M}_{size}$ as the primary comparison index and use $\mathcal{M}_{local-dist}$ as the secondary comparison index). Intuitively, we expect these two measures are much better than size measure alone since size measure is too coarse-grained to distinguish all interesting explanations. The results of the combinations are show in line 7 - 8 of Table 1. It turns out that their combinations are better than any individual interesting measures. It is worth pointing out that these are two very preliminary combinations, and we can definitely further improve the combinations using machine learning techniques. While we believe the current results are sufficient to demonstrate the idea and we leave the detailed study as future work.

**Summary:** When restricted to individual measures, distributional measures achieves the best effectiveness. The combination of structure based measures (e.g., size) with aggregated and distribution-based measures provide better ranking results than any individual measures. To achieve best effectiveness, machine learning algorithms can be used to train best combination of all measures; when efficiency is also a concern, we can restrict the combination on anti-monotonic measures (e.g., $\mathcal{M}_{size}$, $\mathcal{M}_{monocount}$), which will still achieve reasonable effectiveness while can be computed efficiently.

### 5.4.2 Comparing Path and Non-Path Explanations

Based on the user study of previous section, for each target entities pairs, we can pick up to 10 most interesting explanations[14] based on user judgment. Among all top-5 explanations, only 36% of them are paths (64% are non-paths); among all top-10 explanations, 38% of them are paths. The results demonstrate of necessity of including non-paths in the explanation definition.

## 6. RELATED WORK

There are a few recent studies on discovering relationships between various web artifacts. E.g., [20] connects two search terms by extracting pairs of pages based on their common search results; [23] extracts a chain of news articles that connect two news articles based on shared words. Our work is complementary to these as we study entities specifically and leverage a rich knowledge base and a comprehensive set of interestingness measures based on both aggregates and distributions.

Our work is related to the vast literature on keyword search in relational and semi-structured databases [1, 2, 3, 5, 17, 12, 13, 14, 21, 24, 29, 15]. The two major distinctions between *REX* and these works are: (1) We consider connection structures that are more complex than trees and paths for explaining two entities; (2) We introduce two novel families of pattern level interestingness measures.

Our path (instance and pattern) enumeration component can be viewed as a special case of keyword search in databases, where input keywords match exactly two entities. Therefore we can directly adapt algorithms from these works. The first algorithm *PathEnumBasic* is adapted from *BANKS* [5], which does concurrent shortest path run from each target node. The same intuition also comes from *Discover* [14] if we are considering pattern level search. The restriction of "small" relation and the evaluation ordering based on candidate network sharing "frequency" leads us to a very similar solution in our problem settings. The second path enumeration algorithm *PathEnumPrioritized* with node activation score is adapted from *BANKS2* [17]. If we consider pattern level enumeration, the same intuition can also come from *Discover* [14] when we assume

---

[11] During the selection, we removed any pairs with at least one node not recognized by the authors to ensure respondents can easily judge the correctness and interestingness of the explanations.

[12] Discounted cumulative gain is a frequently used ranking measure in web search [16].

[13] The effects of the exact weight values do not change our results much as long as the relative orders are maintained.

[14] We also require the average score of an explanation to be at least 1 to avoid include uninteresting explanations

$b > 0$ in the cost model (i.e., considering the estimated size of join results) when prioritizing the candidate network evaluation.

We emphasize that path enumeration algorithms are not our primary contribution and our framework is flexible enough to take advantage any state-of-art keyword search or path enumeration algorithms. Other related work directly dealing with path enumeration can be found in [28, 19, 7], although they either work in slightly different problem settings or provide similar intuitions as discussed above.

A lot of keyword search papers also discuss ranking based on various interestingness measures. Most of the papers focus on the interestingness at the instance level. Usually, size of the connecting structure is used as the basic metrics. Other enhancements include taking into consideration edge weights [5, 17], node weights [3] and keyword to structure mapping scores [13, 21] inspired by IR techniques. The interestingness measures we proposed are orthogonal to these instance level interestingness measures. We capture the pattern level interestingness by properly aggregating (e.g., count based measures) and normalizing (distributional measures) the instance level measures. Indeed, some work has also considered pattern-level interestingness [24, 29]. However, their problem settings are different: They assume the user of the system to be a domain expert or have a clear search intension (although lack knowledge of the schema or format of data sources). Therefore, these works mainly rely on user feedback to refine and discover the best queries.

There are also quite a few papers on graph mining that mine connecting structures between a set of nodes [8, 10, 18, 22, 25]. However, these algorithms only return a single large connection graph containing a lot of interesting facts, without distilling individual explanations from the remaining part of the connection graph. *REX*, other the other hand, finds multiple interesting explanations and ranks them to describe different aspects of a relationship.

Our work is also closely related to various studies in the frequent graph mining literature. In particular, [9, 26, 27] describes efficient algorithms for identifying frequent sub-graphs from a database of many graphs. While our pruning techniques for anti-monotonic measures are inspired by these algorithms, our problem setting is fundamentally different from their transactional setting: we are mining interesting patterns from a single large graph (i.e., the knowledge base) instead of a database of (relatively) small graphs. More recently, [6] studies the notion of *pattern frequency* in a single graph setting and proposes the notion of monocount as the minimum number of distinct nodes in the original graph that any node in the pattern maps to. Our $\mathcal{M}_{monocount}$ is an extension of this notion. It is worth noting that none of those prior works study distribution-based measures for interestingness.

## 7. CONCLUSION

Given the increasing importance of features like "related searches" on major search engines particularly for entity searches, it is desirable to explain to the users why a given pair of entities are related. And, as far as we know, our work is the first to propose this relationship explanation problem. Furthermore, we studied the desirable properties of relationship explanations given a knowledge base, and formalized both aggregate-based and distribution-based interestingness measures for ranking explanations. The overall problem was decomposed into two sub-problems: *explanation enumeration* and *explanation ranking*; we designed and implemented efficient and scalable algorithms for solving both sub-problems. Extensive experiments with real data show that *REX* discovers high quality explanations efficiently over a real world knowledge base.